\documentclass[aps,showpacs,preprintnumbers,amsmath, amssymb]{revtex4}

\usepackage{float}
\usepackage{graphics,epsfig}
\usepackage{graphicx}
\usepackage{dcolumn}
\usepackage{bm}
\usepackage{mathrsfs}

\begin{document}
\baselineskip=0.8 cm

\title{{\bf The extreme orbital period in scalar hairy kerr black holes}}
\author{Yan Peng$^{1}$\footnote{yanpengphy@163.com}}
\affiliation{\\$^{1}$ School of Mathematical Sciences, Qufu Normal University, Qufu, Shandong 273165, China}

\vspace*{0.2cm}
\begin{abstract}
\baselineskip=0.6 cm
\begin{center}
{\bf Abstract}
\end{center}

In a very interesting paper, Hod has proven that the equatorial
null circular geodesic provides the extreme orbital period to circle a kerr black hole,
which is closely related to the Fermat's principle.
In the present paper, we extend the discussion to
kerr black holes with scalar field hair. We show
that the circle with the extreme orbital period
is still identical to the null circular geodesic.
Our analysis also implies that the Hod's theorem
may be a general property in any axially symmetric
spacetime with reflection symmetry on the equatorial
plane.

\end{abstract}

\maketitle
\newpage
\vspace*{0.2cm}

\section{Introduction}

According to the general relativity, there may be null geodesics outside compact objects,
such as black holes and regular ultra-compact stars \cite{c1,c2}.
The null geodesics can reveal significant features of the curved spacetime geometry.
In particular, the circular null geodesics provide a path for
the massless field to circle compact objects.
Due to it's important applications in astrophysics and theories,
the circular null geodesics have attracted a lot of
attentions \cite{c3,c4,c5}.

The circular null geodesics play an important role in the physics of compact objects.
In particular, the circular null geodesic is closely related to
strong gravitational effects, such as the lensing,
shadow, as well as the gravitational waves \cite{c6,c7,c8,c9,c10,c11,c12,c13,c14,c15,c16}.
From the theoretical aspects, it was found that the null circular orbit
is useful in describing hair distributions outside hairy black holes \cite{s1,s2,s3,s4,s5,s6}.
Especially, there are unstable and stable circular null geodesics around the compact objects.
It was suggested that the characteristic resonances of black holes can be
interpreted as null particles trapped at the unstable circular orbit
and slowly leaking out \cite{r1,r2,r3,r4,r5,r6,r7,r8}.
In the regular ultra-compact star spacetime, the existence of
stable circular null geodesics could trigger
nonlinear instabilities due to that massless fields
can pile up on the stable null obit \cite{us1,us2,us3,us4,us5,us6,us7,us8}.

An important physical problem is to search for
the extreme orbital period to orbit a compact objects \cite{ST1,ST2}.
It should be pointed out that the circular
orbit with the shortest orbital period is distinct from the circular orbit
with the smallest radial paramater due to gravitational red-shift effect.
We should also consider the dragging of inertial frames by spinning compact objects.
Considering the influences of these two interesting physical effects,
Hod showed that the null circular geodesic of a black hole spacetime
is characterized by the extreme orbital period as
measured by asymptotic observers \cite{ST1}.
This property is closely related to the Fermat's principle
in flat spacetime that light propagates along the null trajectories
of extreme time \cite{ST2}.
On the other side, Herdeiro and Radu constructed novel
Kerr black holes with scalar hair \cite{HK1,HK2},
which are equilibrium states and may play an important role in
realistic astrophysical processes \cite{HK3,HK4,HK5,HK6,HK7,HK8}.
In this work, we plan to extend the discussion of orbital
period in \cite{ST1,ST2} to kerr black holes
with scalar field hair.

In the next section, we firstly introduce the kerr black hole with scalar field hair.
Then we investigate the relation between the null circular geodesic
and the extreme orbital period circle.
And the last section contains our main conclusions.

\section{The extreme orbital period in kerr black holes}

Recently, Hod has investigated the fast circle in the background of a probing kerr black hole \cite{ST1,ST2}.
We extensively analyze circular trajectories in the background of
Kerr black holes with scalar field hair.
The asymptotically flat four dimensional deformed scalar hairy Kerr black hole is \cite{HK1,HK2}
\begin{eqnarray}\label{AdSBH}
ds^{2}&=&e^{2F_{1}}(\frac{dr^2}{N}+r^2d\theta^2)+e^{2F_{2}}r^2\sin^2\theta(d\phi-Wdt)^{2}-e^{2F_{0}}Ndt^{2},
\end{eqnarray}
where $F_{0}$, $F_{1}$, $F_{2}$ and W are functions of the radial coordinate r.
And N can be expressed as $N=1-\frac{r_{H}}{r}$ with $r_{H}$ as the event horizon.
Since the spacetime is asymptotically flat,
the functions are characterized by $F_{0}\rightarrow 0, F_{1}\rightarrow 0, F_{2}\rightarrow 0, W\rightarrow 0$
as approaching the infinity.
We take the usual angular coordinates $\theta\in [0,\pi]$ and $\phi\in [0,2\pi]$.
The equatorial plane of the black hole is characterized by $\theta=\frac{\pi}{2}$.
With reflection symmetry, the circular orbit lies on the equatorial
plane \cite{HK2}.

And we take the ansatz of the scalar field in the form \cite{HK1,HK2}
\begin{eqnarray}\label{AdSBH}
\mathcal{\psi}&=R(r,\theta)e^{i(m\phi-\omega t)},
\end{eqnarray}
where $\omega$ is the frequency of the scalar field and $m=\pm1,\pm2...$ is the azimuthal harmonic index.

Now we would like to search for the circular trajectory with the radius $r=r_{extreme}$,
which corresponds to the extreme orbital period measured by asymptotic observers.
We shall consider circular orbits in the black hole equatorial plane
characterized by $\theta=\frac{\pi}{2}$.
In order to minimize the orbital period for a given
radius r, one should move as close as possible to the speed of light.
In this case, the orbital period can be obtained from Eq. (1) with
$ds=dr=d\theta=0$ and $\Delta\phi=2\pi$ \cite{ST1,ST2}:
\begin{eqnarray}\label{AdSBH}
T(r)=-\frac{2\pi(g_{t\phi}\pm\sqrt{g_{t\phi}^2-g_{tt}g_{\phi\phi}})}{g_{tt}}.
\end{eqnarray}
Another case of $\Delta\phi=-2\pi$ can be obtained by the transformation
$g_{t\phi}\rightarrow -g_{t\phi},~\phi\rightarrow-\phi$.
The circular trajectory around the central black hole
with the extreme orbital period is characterized by
\begin{eqnarray}\label{AdSBH}
T'(r=r_{extreme})=0,
\end{eqnarray}
where the prime is a derivative with respect to the coordinate r.
This yields the characteristic equation
\begin{eqnarray}\label{AdSBH}
G(r_{extreme})=T'(r_{extreme})=(g_{t\phi}\pm\sqrt{g_{t\phi}^2-g_{tt}g_{\phi\phi}})\frac{g'_{tt}}{g^{2}_{tt}}-(g'_{t\phi}\mp \frac{g_{tt}g'_{\phi\phi}-2g_{t\phi}g'_{t\phi}+g'_{tt}g_{\phi\phi}}{2\sqrt{g_{t\phi}^2-g_{tt}g_{\phi\phi}}})\frac{1}{g_{tt}}=0.
\end{eqnarray}

Now we derive the relevant equations of the null circular geodesic radius
$r=r_{\gamma}$ \cite{c2,c3,c4}.
The Lagrangian describing the geodesics in the spacetime (1) is given by
\begin{eqnarray}\label{BHg}
2\mathcal{L}=g_{tt}\dot{t}^2+2g_{t\phi}\dot{t}\dot{\phi}+g_{rr}\dot{r}^2+g_{\phi\phi}\dot{\phi}^2,
\end{eqnarray}
where a dot denotes the ordinary differentiation with respect to the affine parameter along the geodesic.

Since the metric has the time Killing vectors $\partial t$
and the axial Killing vector $\partial\phi$, there are two
constants of motion labeled as E and L.
The generalized momenta can be
derived from the Lagrangian as
\begin{eqnarray}\label{BHg}
p_{t}=g_{tt}\dot{t}+g_{t\phi}\dot{\phi}=-E=const,
\end{eqnarray}
\begin{eqnarray}\label{BHg}
p_{\phi}=g_{t\phi}\dot{t}+g_{\phi\phi}\dot{\phi}=L=const,
\end{eqnarray}
\begin{eqnarray}\label{BHg}
p_{r}=g_{rr}\dot{r}~.
\end{eqnarray}
The Hamiltonian can be expressed as
$\mathcal{H}=p_{t}\dot{t}+p_{r}\dot{r}+p_{\phi}\dot{\phi}-\mathcal{L}$,
which implies
\begin{eqnarray}\label{BHg}
2\mathcal{H}=-E\dot{t}+L\dot{\phi}+g_{rr}\dot{r}^2=\delta=const.
\end{eqnarray}
Here we can take the value $\delta=0$ in the case of null geodesics.

According to (10), we arrive at the relation
\begin{eqnarray}\label{BHg}
\dot{r}^2=\frac{1}{g_{rr}}[E\dot{t}-L\dot{\phi}]
\end{eqnarray}
for null geodesics.

From relations (7) and (8), we easily get $\dot{t}$ and $\dot{\phi}$ in the form
\begin{eqnarray}\label{BHg}
\dot{t}=\frac{Eg_{\phi\phi}+Lg_{t\phi}}{g_{t\phi}^2-g_{tt}g_{\phi\phi}},~~~~~~~~~~\dot{\phi}=-\frac{Eg_{t\phi}+Lg_{tt}}{g_{t\phi}^2-g_{tt}g_{\phi\phi}}.
\end{eqnarray}

Substituting (12) into (11), there is
\begin{eqnarray}\label{BHg}
\dot{r}^2=\frac{E^2}{g_{rr}}[\frac{g_{\phi\phi}+2bg_{t\phi}+b^2g_{tt}}{g_{t\phi}^2-g_{tt}g_{\phi\phi}}],
\end{eqnarray}
where we introduce a new constant $b=L/E$.

The requirement $\dot{r}^2=0$ for a null circular geodesic yields
\begin{eqnarray}\label{BHg}
b_{\pm}=-\frac{g_{t\phi}\pm\sqrt{g_{t\phi}^2-g_{tt}g_{\phi\phi}}}{g_{tt}}.
\end{eqnarray}

The requirement $(\dot{r}^2)'=0$ and relation (14) yield the equation
\begin{eqnarray}\label{BHg}
\frac{G(r_{\gamma})}{g_{rr}\sqrt{g_{t\phi}^2-g_{tt}g_{\phi\phi}}}=0.
\end{eqnarray}

In the probe Kerr black hole spacetime, Hod proved that the extreme
period circle radius equation and the null circular geodesics
equation share the same roots though they seem to be different
in the form, see (21) and (31) in \cite{ST1}.
Here we show that the shortest period circle radius equation (5)
is in fact the same as the null circular geodesic radius equation (15)
except the product factor $\frac{1}{g_{rr}\sqrt{g_{t\phi}^2-g_{tt}g_{\phi\phi}}}$.
And the similarity between (5) and (15) implies that the Hod's theorem
may be a general property in any axially symmetric spacetime
with reflection symmetry on the equatorial
plane. In the scalar hairy kerr black hole,
we can prove that (5) and (15) share the same roots by
showing $(g_{rr}\sqrt{g_{t\phi}^2-g_{tt}g_{\phi\phi}})|_{r=r_{\gamma}}\neq0$.

The physical extreme period circular trajectories cannot be
on the horizon ($r_{extreme}\neq r_{H}$) since the horizon will
absorb all matter fields. So we only focus on the null circular geodesic radius $r_{\gamma}> r_{H}$.
It may be interesting to extend our analysis to horizonless hairy compact stars \cite{star-1}-\cite{star-8}.
In this work, we only focus on the spacetime with a horizon.
At the null circular geodesic radius $r_{\gamma}$ of hairy Kerr black hole, there is
\begin{eqnarray}\label{BHg}
N(r_{\gamma})> 0,
\end{eqnarray}
\begin{eqnarray}\label{BHg}
g_{rr}(r_{\gamma})=\frac{e^{2F_{1}}}{N}> 0,
\end{eqnarray}
\begin{eqnarray}\label{BHg}
g^2_{t\phi}-g_{tt}g_{\phi\phi}=r^2N(r_{\gamma})e^{2F_{0}+2F_{2}}> 0.
\end{eqnarray}

According to (17) and (18), there is the relation
\begin{eqnarray}\label{BHg}
(g_{rr}\sqrt{g_{t\phi}^2-g_{tt}g_{\phi\phi}})|_{r=r_{\gamma}}\neq0.
\end{eqnarray}
Now we can conclude that the null circular
geodesic equation (15) shares the same roots
with the extreme period characteristic relation (5).
Thus it means that the null circular geodesic of hairy Kerr black hole provides the path with
the extreme orbital period as measured by asymptotic observers
\begin{eqnarray}\label{BHg}
r_{extreme}=r_{\gamma}~.
\end{eqnarray}
We point out that Eqs. (5) and (15)
may have various solutions. In such cases,
extreme period circular trajectories
still correspond to null circular geodesics.
It should be emphasized that apart from the general ansatz (1)
and the qualitative relations (16)-(18), nothing specific of the Kerr black holes
with scalar hair solutions is used. So the relation (20) holds for
any axi-symmetric, stationary spacetime that takes the
generic form (1) and obeys (19).

\section{Conclusions}

We investigated the equatorial circular orbits
in backgrounds of kerr black holes with scalar field hair.
In this backreacted kerr black hole,
we showed that the extreme period circle radius equation
is similar to the null circular geodesic radius equation.
We further found that null circular
geodesics provide extreme period equatorial circular
trajectories. Our analysis also implies that the Hod's theorem
may be a general property in any axially symmetric
spacetime with reflection symmetry on the equatorial
plane. These results are in analogy with
the Fermat's principle in the flat spacetime,
which asserts that light takes the path
with the extreme traveling time.

\begin{acknowledgments}

We would like to thank the anonymous referee for the constructive suggestions to improve the manuscript.
This work was supported by the Shandong Provincial Natural Science Foundation of China under Grant
No. ZR2018QA008.

\end{acknowledgments}

\end{document}